# Arrhythmia Detection using Mutual Information-Based Integration

Othman Mohammad Soufan* and Samer Arafat**

*Math and Computer Sciences and Engineering Department,
King Abdullah University of Science and Technology,
Thuwal, Saudi Arabia
**othman.soufan@kaust.edu.sa**

**ICS Department
King Fahd University of Petroleum and Minerals,
Dahran, Saudi Arabia
**arafat@kfupm.edu.sa**

*Abstract*— **The aim of this paper is to propose an application of mutual information-based ensemble methods to the analysis and classification of heart beats associated with different types of Arrhythmia. Models of multilayer perceptrons, support vector machines, and radial basis function neural networks were trained and tested using the MIT-BIH arrhythmia database. This research brings a focus to an ensemble method that, to our knowledge, is a novel application in the area of ECG Arrhythmia detection. The proposed classifier ensemble method showed improved performance, relative to either majority voting classifier integration or to individual classifier performance. The overall ensemble accuracy was 98.25%.**

## I. Introduction

At the top of the right chamber of the human heart, an electrical signal is generated from the Sino Atrial node which stimulates the heartbeat [1]. The heart may experience abnormal increase or decrease in its beat rate which is known as arrhythmia [2]. In order to detect this type of abnormality, an electrocardiogram (ECG) device that measures the variations in the electrical signals of heart is used.

As reported by the American Heart Association (AHA) "Each year about 295,000 emergency medical services-treated out-of-hospital cardiac arrests occur in the United States" [2]. Thus, having an automated system that is able to diagnose heart beats and offer an early detection of arrhythmia would greatly help in preventing cardiac arrests, thus saving people who might face such abnormalities. Also, it can help cardiologists in monitoring the heart beat rates and deciding on the specific types of arrhythmia.

The aim of this research is to enhance Arrhythmia analysis and detection ratios by combining different classifiers into one model. This analysis is devoted to decisions concerned with recognizing and classifying the patterns of the abnormal electrical signals of the heart beats. This work combines different types of classifiers and compares their performance as individual models and as integrated ensembles. The ensemble models were integrated according to majority voting or normalized mutual information-weighted sum rules. Individual classifier models include back-propagation- trained multilayer perceptrons, support vector machines and radial basis function neural networks. The simulation results were based on training and testing the extracted samples of the well-known MIT-BIH arrhythmia database.

This paper is organized as follows. In Section II, we briefly review works related to ECG beat detection and classification. Our implemented method for integrating different classification models is described in Section III. Data used in this research and extracted classification features are presented in Sections IV and V, respectively. In Section VI, the evaluation results of the different suggested classification models are reported. Finally, conclusions are made in Section VII.

## II. Related Work

Different recognition and classification models dedicated for ECG arrhythmia detection were proposed in the literature. Considering artificial neural networks, several projects that are based on applying multilayer perceptron (MLP) neural networks were developed [3-8]. Ebrahimzadeh *et al.*. [3] investigated different designs of neural networks and experimented with nine different learning algorithms on the data of the MIT-BIH database. Another method based on MLP was proposed by Niwas *et al.* [5] in which an overall accuracy of 99.02% was achieved for nine different classes of arrhythmia. However, the accuracy was achieved by training a large number of datasets ranging from 15 to 40 datasets per class. Each data set is supposed to contain at least 1500 sample beats. Also, Zhu *et al.* [6] examined three models of binary MLP classification and achieved accuracy above 95% with a small number of training examples. In the study of Zeybekoğ *et al.* [8], five types of ECG signals including Normal ECG, Ventricular Tachycardia, Left Bundle Branch Block, Right Bundle Branch Block and Atrial Fibrillation were classified using MLP. They achieved a general accuracy of 82% in their classification.

For radial basis function (RBF) neural network, different studies achieved high classification results for multiclass recognition. In the study of Guangying *et al.* [9], they were able to reach a classification rate of 90% for the normal class and paced heart beats class. Moreover, a specificity of 97.42% and sensitivity of 88.37% were recognized for classifying six types of beats including normal beat, premature ventricular contraction (PVC), fusion of ventricular and normal beat (F), atrial premature beat (APB), right bundle branch block beat (R), and fusion of paced and normal beat (f) of samples taken from the MIT-BIH database [10].

Different studies that are based on support vector machines (SVM) were held to detect arrhythmia in ECG signals [11-14]. A comparison between one-against-one and one-against-all multiclass SVM classification were performed and the one-against-all produced the best results regarding the case of arrhythmia detection[11]. In Joshi *et al.* [14], a hybrid SVM system was implemented to classify different types of heart beat arrhythmias. The hybrid SVM model is composed of two multiclass SVMs and a binary SVM one each dedicated to different classification task. In this model an overall validation accuracy of 90.89% was achieved on the MIT-BIH data [14].

Osowski *et al.* presented the idea of integrating multiple classifiers including MLP, SVM and hybrid fuzzy network [15]. The implemented model was able to reduce the misclassification error by more than 20%. In the study of Osowski *et al.*, the data size used for training and testing were 5140 and 5178 data pairs respectively.

### III. IMPLEMENTED METHODS

In this research, three classifier models were used to predict different types of arrhythmia. The models include the back-propagation trained Multilayer Perceptron (BP MLP) version of artificial neural networks, support vector machines (SVM) and radial basis function neural networks (RBF NN). The classifications of these models were then integrated into a single model using two approaches that will be discussed in Part D of this Section.

*A. Back-Propagation Artificial Neural Networks*

The general BP MLP training algorithm is based on input forward propagation and error backward propagation followed by an update on the weights of the network using gradient methods. Based on the results of Ebrahimzadeh *et al.* experiments [3], the Levenberg-Marquardt LM training algorithm was implemented in our model with 35 hidden units for the first hidden layer. Also, the sigmoid activation function was chosen for the hidden layer with a linear one on the output layer.

*B. Radial Basis Function Neural Networks*

In RBF NN, the square of distances between the input vector $x$ and mean vector $c_m$ for the radial basis functions are computed using $\|x - c_m\|^2$ and the output of the m-th hidden unit is represented by $y_m$ as in (2).

$$y_m = \exp\left[\frac{-\|x - c_m\|^2}{2\sigma^2}\right] \quad (2)$$

Next, the output values of the hidden units are weighted and summed to produce the final results [17]. In (2), $\sigma$ is the spread factor interpreted as the radius of the radial basis function. The spread parameter plays a major rule in classifying the data and a spread factor of 105 was selected experimentally to classify the ECG signals.

*C. Support Vector Machines*

SVM is an efficient classification algorithm that project non-linearly separable data into higher-dimensional space using kernel techniques [18]. In higher dimensions, data points can become linearly separable. Also, it computes a margin that maximally separates classes in the space.

In this research, a polynomial kernel $K(x,w)$ of degree two was applied to classify the normal beats and four types of arrhythmias. In (3), the input training vectors ($x$) and weights vectors ($w$) and the bias or cost parameter ($b_0$) constitute the terms for the applied polynomial kernel.

$$K(x,w) = (x^T w + b_0)^2 \quad (3)$$

In order to experiment with SVM using the ECG data, LIBSVM library was used [19]. Several kernels were tested and a polynomial kernel of degree 2 was finally implemented.

*D. Integration Method*

Combining different classification models is an objective that has shown improved performance in different studies [20].

Lei Xu *et al.* discuss three types of outputs that are combined according to different approaches [20]. One type is based on the majority voting principle in which the output of the classification is of integer values representing the class the sample belongs to. Another type ranks candidate labels for which a candidate subset combining and re-ranking approach can be applied. The third type of output is a continuous real valued output viewed as a confidence value between [0, 1].

In our work, we generate two types of outputs; an integer output declaring class labels as {0, 1} and another real continuous output which can be regarded as a confidence value between [0, 1]. For the first case, we set three thresholds on the output of the three classification models and, in turn, apply a simple majority voting with no weighting factor. This computes the first integrated ensemble model.

Regarding the case of continuous real-valued outputs we implemented a fusion technique that integrates the 3 classifier model output values, in the following manner. First, we normalize output vectors of all three models, relative to their actual (computed) classes, using the corresponding second norms. In other words, the actual output vectors ($a_{ij}$) for the i[th] classifier and j[th] class, are normalized using their corresponding second norm values. Next, normalized output vectors are integrated according to the following proposed weighted sum rule:

$$S_j = n_{1j} \frac{a_{1j}}{\|a_{1j}\|} + n_{2j} \frac{a_{2j}}{\|a_{2j}\|} + n_{3j} \frac{a_{3j}}{\|a_{3j}\|} \quad (5)$$

The weights ($n_{ij}$) in (5) are computed using the normalized mutual information measure $I_{norm}(a_{ij}, d_{ij})$ described in (6). In (6), $I(a_{ij}, d_{ij})$ denotes the mutual information between the actual data $a_{ij}$ coming from the i[th] classifier for the j[th] class and the desired (true) prediction ($d_{ij}$) for the same classifier i and class j [21]. $H(a_{ij})$ is the entropy measure function. The computation of the mutual information depends on the number of true positives (TP), number of true negatives (TN), number of false positives (FP) and number of false negatives (FN). The normalized mutual information was used as a weight rather than using the sensitivity or specificity because it considers TP, TN, FP and FN it its formula while both sensitivity or specificity consider only two of them. So, the normalized mutual information is a global measure for the reduction of uncertainty of true class labels when actual class labels are observed (i.e, computed) while sensitivity and specificity are local ones.

$$n_{ij} = I_{norm}(a_{ij}, d_{ij}) = \frac{I(a_{ij}, d_{ij})}{H(a_{ij})} \quad (6)$$

with

$$H(a_{ij}) = -\frac{TP+FN}{N}\log\left[\frac{TP+FN}{N}\right] - \frac{TN+FP}{N}\log\left[\frac{TN+FP}{N}\right] \quad (7)$$

and

$$\begin{aligned}I(a_{ij}, d_{ij}) = &-H\left(\frac{TP}{N}, \frac{TN}{N}, \frac{FP}{N}, \frac{FN}{N}\right) \\ &-\frac{TP}{N}\log\left[\frac{TP+FP}{N}\frac{TP+FN}{N}\right] \\ &-\frac{FN}{N}\log\left[\frac{TP+FN}{N}\frac{TN+FN}{N}\right] \\ &-\frac{FP}{N}\log\left[\frac{TP+FP}{N}\frac{TN+FP}{N}\right] \\ &-\frac{TP}{N}\log\left[\frac{TN+FN}{N}\frac{TN+FP}{N}\right]\end{aligned} \quad (8)$$

for which

$$H\left(\frac{TP}{N}, \frac{TN}{N}, \frac{FP}{N}, \frac{FN}{N}\right) = -\frac{TP}{N}\log\frac{TP}{N} - \frac{TN}{N}\log\frac{TN}{N} - \frac{FP}{N}\log\frac{FP}{N} - \frac{FN}{N}\log\frac{FN}{N} \quad (9)$$

Finally, a threshold is determined, experimentally, for the results of the weighted sum ensemble method to get the {0, 1} class labels and performance measures for classifying five classes are computed and compared to other individual models.

I. EXPERIMENTAL DATA

Between 1975 and 1979, 48 annotated ECG records that belong to 47 subjects were analyzed by the BIH Hospital in Boston. Nearly 60% of the records were taken from inpatients. Also, 23 records in the built database were chosen randomly from a set of over 4000 24-hour ECG tapes. The other 25 records were chosen to include specific types of arrhythmias [22].

Each record represents about 30 minutes of the signal sampled at 360 Hz. The database consists in total of roughly 109,000 beats that were manually annotated by at least two cardiologists [22].

In our research, the segmentation for the QRS wave and waveform boundary intervals was done using an algorithm implementation that is available as part of the *ecgpuwave PhysioToolkit* software [26]. The part of the function that segments the QRS complex is based on the algorithm of Pan and Tompkins [27] with some modification to exploit the information of the slope [28]. For the end points recognition of the P and T-waves, an algorithm described in [29] and evaluated in [29] and [30] was used as a component of the segmentation function.

Our preprocessing of the data included a check for missing segmentation values and/or missing class labels. The available software described, above, segmented the data with high accuracy but has sometimes made an error in the identification of the QRS segment. This resulted in either missing data values or missing labels due to the absence of proper R wave presence in the final segmentation. Beats corresponding to such missing information were excluded. Due to this preprocessing, out of about 109,400 total beats, we used 108,232 beats, or 98.6% of the original data

We aimed at training using the following classes: the normal beats and four types of arrhythmia. These arrhythmias include Premature Ventricular Contraction PVC, Atrial Premature Beat APB, Right Bundle Branch Block RBBB and Left Bundle Branch Block LBBB types of arrhythmia. The distribution of the number of samples for each class is shown in TABLE 1. A random 4% sample size of the total 108,232 samples was picked for training. The remaining 96% data were used for testing.

TABLE 1 Summary of Extracted Data from MIT-BIH Database

| Class | Number of Samples |
| --- | --- |
| Normal | 74385 |
| PVC | 6730 |
| APB | 2356 |
| RBBB | 7205 |
| LBBB | 8033 |
| Other | 9523 |

II. EXTRACTED ECG FEATURES

Extracting features and selecting the appropriate ones will affect the prediction performance, help in producing faster and more cost-effective predictors [23]. In order to solve a feature selection problem, a set of questions might be asked as suggest by Guyon and Elisseeff in [23]. The first proposed question is about deriving set of "ad hoc" features based on the knowledge of the domain being examined. In ECG research, a considerable set of clinical features can be established from the ECG domain knowledge.

Considering the domain of ECG, the standard clinical ECG features are generally the measurement of inter-beat timings and amplitudes [24]. Some of the major features that can help in detecting arrhythmia and differentiate between normal and abnormal beats are PR interval, QRS width and the corrected QT interval of the ECG signal [24][25]. Also, features such as P-wave amplitude, T-wave amplitude and QRS height affect the decision on the beat type [24]. Some of these features are illustrated in Fig. 1.

The time-domain diagnostic and morphologic features of our research include the widths of the PR, QRS, QT and RR intervals. The other extracted features include the amplitude of the QRS interval, the mean and standard deviation of the amplitudes in the range of the QRS complex, the QT interval and the RR interval.

III. RESULTS

Three classifiers and two integration methods were applied for classifying five classes in which one class is normal and four are abnormal. Sensitivity and specificity were considered as performance measures in comparing the classification models rather than accuracy since the data is not evenly distributed between classes. The summary of the results are represented in TABLES 1, 2, 3, 4 and 5 in the next page in which each table represents a different classification model.

For the normal class, both of the suggested integrated ensemble methods achieved higher results than the individual ones. If we take the average of sensitivity and specificity, the ensemble method that uses the majority voting rule, achieved the best results with an average sensitivity and specificity of 94.39%.

For the premature ventricular contraction PVC arrhythmia class, the ensemble method that uses the weighted sum rule with a normalized mutual information factor attained the higher results with sensitivity of 91.61% and specificity of 99.32%. On the other hand, with a sensitivity of 72.28% and specificity of 99.39%, the RBF NN outperformed the other methods in detecting atrial premature beats APB.

Considering the right bundle branch block RBBB, the weighted sum ensemble method reached a sensitivity of 93.72% and specificity of 99.77% that are higher than all other methods. In the case of the left bundle branch block LBBB class, the weighted sum ensemble also achieved the highest results with an average sensitivity and specificity of 96.9%.

Overall, TABLE 7 shows 98.25% accuracy, 97.35% specificity, and 89.82% sensitivity for the weighted sum ensemble. These results are either comparable or slightly outperform the majority voting ensemble, as well as the 3 individual classifier performances.

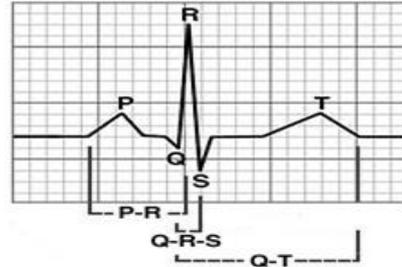

Figure 1. Illustration for PR, QRS, QT Intervals on a Normal ECG Signal [31]

IV. CONCLUSION

We implemented classification techniques that included BP MLP, RBF NN, and SVM. Two different ensemble methods were proposed. One depends on a simple majority voting while the second is based on a weighted sum rule in which the weights are normalized mutual information factors. RBF NN attained the best performance in detecting two types of arrhythmias. Also, the small number of training samples in which 4% random samples were chosen from the MIT-BIH database to train the models while the 96% were used for testing. Thus, models with high performance are ones that have a robust generalization.

Results have demonstrated that a mutual information-based approach slightly outperformed a majority voting ensemble as well as individual classifiers. This demonstrates the significance of ensemble methods, although these methods might not outperform individual classifiers, all the time. Furthermore, these results are comparable to the most recent available works.

The proposed mutual information-based approach brings a focus to an ensemble method that is based on normalized mutual information weighted sum that, to our knowledge, was not applied, so far, in the area of ECG-based Arrhythmia detection. Results have demonstrated this approach's performance, relative to either a majority voting method or individual classifier performance. The mutual information measure computes the reduction of uncertainty of actual output relative to desired output. Therefore, if this measure computes a high value that is multiplied with a corresponding actual class output, then this would enhance that output, relative to other classes output with corresponding lower (computed) values of this measure. This is intuitive because if this measure computes a high value for a certain class then the class's corresponding actual output is highly in agreement with its desired outputs. Also, this is interpreted in the sense that individual classifiers with the highest information content receive relatively higher weights.

For future work, we aim to implement more sophisticated integrating and ensemble methods that can optimize this research's results. Also, we look forward classifying more classes of arrhythmia. Moreover, we aim at applying our suggested models on different well-representative databases.

TABLE 2  BP MLP Results for Classifying 5 Classes of Heartbeats

| Class | Accuracy % | Sensitivity % | Specificity % | False Positive Rate % | False Negative Rate % |
|---|---|---|---|---|---|
| Normal | 95.18% | 97.09% | 90.96% | 9.04% | 2.91% |
| PVC | 98.43% | 91.58% | 98.88% | 1.12% | 8.42% |
| APB | 98.21% | 57.12% | 99.12% | 0.88% | 42.88% |
| RBBB | 99.25% | 93.33% | 99.67% | 0.33% | 6.67% |
| LBBB | 97.92% | 92.61% | 98.35% | 1.65% | 7.39% |

TABLE 3  RBF NN Results for Classifying 5 Classes of Heartbeats

| Class | Accuracy % | Sensitivity % | Specificity % | False Positive Rate % | False Negative Rate % |
|---|---|---|---|---|---|
| Normal | 94.75% | 96.46% | 91.00% | 9.00% | 3.54% |
| PVC | 98.61% | **91.70%** | 99.07% | 0.93% | **8.30%** |
| APB | 98.80% | **72.28%** | 99.39% | 0.61% | **27.72%** |
| RBBB | 99.18% | 92.91% | 99.62% | 0.38% | 7.09% |
| LBBB | 98.27% | 91.22% | 98.83% | 1.17% | 8.78% |

TABLE 4  SVM Results for Classifying 5 Classes of Heartbeats

| Class | Accuracy % | Sensitivity % | Specificity % | False Positive Rate % | False Negative Rate % |
|---|---|---|---|---|---|
| Normal | 92.99% | 95.91% | 86.56% | 13.44% | 4.09% |
| PVC | 98.16% | 82.79% | 99.18% | 0.82% | 17.21% |
| APB | 98.60% | 53.13% | 99.61% | 0.39% | 46.87% |
| RBBB | 98.99% | 91.74% | 99.51% | 0.49% | 8.26% |
| LBBB | 97.61% | 86.59% | 98.49% | 1.51% | 13.41% |

TABLE 5  Majority Voting Integrated Results for Classifying 5 Classes of Heartbeats

| Class | Accuracy % | Sensitivity % | Specificity % | False Positive Rate % | False Negative Rate % |
|---|---|---|---|---|---|
| Normal | 95.51% | 97.38% | **91.39%** | **8.61%** | 2.62% |
| PVC | 98.73% | 91.13% | 99.23% | 0.77% | 8.87% |
| APB | **98.90%** | 61.46% | **99.73%** | **0.27%** | 38.54% |
| RBBB | 99.34% | 93.34% | 99.76% | 0.24% | 6.66% |
| LBBB | **98.59%** | 92.00% | **99.12%** | **0.88%** | 8.00% |

TABLE 6  Weighted Sum Integrated Results for Classifying 5 Classes of Heartbeats

| Class | Accuracy % | Sensitivity % | Specificity % | False Positive Rate % | False Negative Rate % |
|---|---|---|---|---|---|
| Normal | **95.70%** | **98.54%** | 89.43% | 10.57% | **1.46%** |
| PVC | **98.84%** | 91.61% | **99.32%** | **0.68%** | 8.39% |
| APB | 98.86% | 70.29% | 99.50% | 0.50% | 29.71% |
| RBBB | **99.37%** | **93.72%** | **99.77%** | **0.23%** | 6.28% |
| LBBB | 98.46% | **94.96%** | 98.75% | 1.25% | **5.04%** |

TABLE 7  Summary of Performance Results of the Proposed ECG Classification Models

| Class | BP MLP | RBF NN | SVM | Majority Voting Ensemble | Mutual Information Weighted Ensemble |
|---|---|---|---|---|---|
| Accuracy % | 97.80% | 97.92% | 97.27% | 98.21% | **98.25%** |
| Sensitivity % | 86.35% | 88.92% | 82.03% | 87.06% | **89.82%** |
| Specificity % | 97.40% | 97.58% | 96.67% | **97.85%** | 97.35% |

ACKNOWLEGEMENT

The authors would like to acknowledge the support provided by the Deanship of Scientific Research at King Fahd University of Petroleum & Minerals (KFUPM).